\documentclass{aa}

\usepackage{graphicx}
\usepackage{txfonts}
\usepackage{newtxtext}  
\usepackage[varvw]{newtxmath} 
\bibpunct{(}{)}{;}{a}{}{,} 
\usepackage{epstopdf, epsfig}
\usepackage[colorlinks=true,linkcolor=blue,citecolor=blue]{hyperref}
\usepackage{amsmath,amsfonts,amssymb}
\usepackage{txfonts}
\usepackage[utf8]{inputenc}
\usepackage[T1]{fontenc}
\usepackage{color} 
\usepackage{xcolor}
\usepackage{orcidlink}

\newcommand{\hr}{\hat{r}}

\begin{document}

   \title{Neutrinos from stochastic acceleration in black hole environments}
   

   \author{Martin Lemoine\orcidlink{0000-0002-2395-7812}
          \inst{1}
          \and
          Frank Rieger\orcidlink{0000-0003-1334-2993}
          \inst{2,3}
          }
   \institute{
            Astroparticule et Cosmologie (APC), CNRS - Université Paris Cité, 75013 Paris, France\\ 
            \email{mlemoine@apc.in2p3.fr}
            \and
            Max Planck Institute for Plasma Physics (IPP), Boltzmannstraße 2, 85748 Garching, Germany
            \and
            Institute for Theoretical Physics, Heidelberg University, Philosophenweg 12, 69120 Heidelberg, Germany\\
            \email{frank.rieger@ipp.mpg.de}
             }


 
  \abstract 
  {Recent experimental results from the IceCube detector and their phenomenological interpretation suggest that the magnetized turbulent corona of nearby X-ray luminous Seyfert galaxies can produce $\sim 1-10\,$TeV neutrinos via photo-hadronic interactions. We investigate the physics of stochastic acceleration in these environments in detail and examine the conditions under which the inferred proton spectrum can be explained. To this end, we used recent findings on particle acceleration in turbulence and paid particular attention to the transport equation, notably for transport in momentum space, turbulent transport outside of the corona, and advection through the corona. We first remark that the spectra we obtained are highly sensitive to the value of the acceleration rate, for instance, to the Alfvénic velocity. Then, we examined three prototype scenarios, one scenario of turbulent acceleration in the test-particle picture, another scenario in which particles were preaccelerated by turbulence and further energized by shear acceleration, and a final scenario in which we considered the effect of particle backreaction on the turbulence (damping), which self-regulates the acceleration process. We show that it is possible to obtain satisfactory fits to the inferred proton spectrum in all three cases, but we stress that in the first two scenarios, the energy content in suprathermal protons has to be fixed in an ad hoc manner to match the inferred spectrum at an energy density close to that contained in the turbulence. Interestingly, self-regulated acceleration by turbulence damping naturally brings the suprathermal particle energy content close to that of the turbulence and allowed us to reproduce the inferred flux level without additional fine-tuning. We also suggest that based on the strong sensitivity of the highest proton energy to the Alfvénic velocity (or acceleration rate), any variation in this quantity in the corona might affect (and in fact, set) the slope of the high-energy proton spectrum. }


   \maketitle
   
\section{Introduction}
Supermassive black holes in active galactic nuclei (AGNs) have long been regarded as potential accelerators to extreme energies~\citep[e.g.,][]{1979ApJ...232..106E,1981ICRC....1..238B,1992PhRvL..69.2885P}, whether in the accretion disk itself~\citep[e.g.,][]{1996ApJ...456..106D,2005ApJ...619..306A,14Lynn,2019MNRAS.485..163K,2019PhRvD.100h3014K},
in the magnetosphere~\citep[e.g.,][]{2002APh....16..265L,2002PhRvD..66b3005A}, around the base of the jet~\citep[e.g.,][]{2000A&A...353..473R,2002A&A...396..833R,2009A&A...506L..41R}, or more recently, through tidal disruption events~\cite[e.g.,][]{2014arXiv1411.0704F,2018A&A...616A.179G,2020ApJ...902..108M,2023ApJ...948...42W}. 

The recent announcement the arrival directions of (a subset of) high-energy ($\epsilon_\nu\gtrsim 1-10$ TeV) neutrinos are correlated with local AGNs, in particular, from the nearby ($d\sim 10$ Mpc) Seyfert galaxy NGC 1068~\citep{2022Sci...378..538I,2023arXiv230507086H,2025ApJ...981..131A}, has placed these theoretical considerations on a solid footing, triggered a surge of interest in these systems, and more generally, paved the way for a novel view of the origin of high-energy neutrinos~\citep{2024PhRvL.132j1002N,Padovani2024}. If it is confirmed, the detection of these neutrinos, with a flux higher by an order of magnitude than the gamma-ray flux that would be expected in association, suggests that they are produced through hadronic interactions of protons with energies $\epsilon_p\gtrsim 30\,$TeV in compact environments, where gamma-gamma absorption through pair production becomes high enough to limit the outgoing gamma-ray flux~\citep{2020PhRvL.125a1101M}. Particle acceleration can be envisaged at accretion shocks, in the magnetized turbulence, through magnetic reconnection, or through a combination of these processes~\citep{2020PhRvL.125a1101M,2020ApJ...891L..33I,2021ApJ...922...45K,2022ApJ...941L..17M,2022ApJ...939...43E,2023ApJ...956....8F,2024ApJ...961L..14F,2024ApJ...961L..34M,2024PhRvD.109j1306M,2024ApJ...974...75F,2024ApJ...972...44D,2024JCAP...09..075A}. In principle, the corona is strongly magnetized and turbulent, and stochastic acceleration therefore emerges as a prime candidate. The main uncertainty on the expected ion and secondary neutrino spectra in this context stems from the modeling of particle 
acceleration and from the description of the corona itself.

Several authors have recently pointed out differences between the magnitude of the neutrino flux inferred from NGC~1068 and that expected from the overall population \citep[i.e., NGC 1068 should be significantly more luminous than other Seyferts, see][]{Waxman24} and with the available energy budget of the NGC~1068 corona itself~\citep{2024PASJ...76..996I}. Pending confirmation of the  IceCube results, which reach the $4.2\sigma$ confidence level so far, we take these observations at face value and provide a fresh and detailed discussion of stochastic acceleration in a turbulent magnetized corona here in order to discuss the conditions under which the overall flux and observed spectral shape can be recovered. To this end, we followed recent insights into the physics of particle acceleration in turbulent and sheared velocity flows to investigate the acceleration rate in the turbulent corona and at the base of the black hole outflow. We payed particular attention to the transport equation in stochastic acceleration, and we addressed the possible backreaction of accelerated particles on the turbulence and its consequences for phenomenology. As we discuss below, this backreaction, which generically leads to the damping of the turbulence and therefore to the modification of the accelerated particle spectra~\citep[e.g.,][and references therein]{2024PhRvD.109f3006L}, appears likely based on the high ion energy flux inferred from IceCube data. We also examined the possible reacceleration of particles by shear acceleration at the base of the jet, close to the black hole. We have a hierarchical picture in mind in which particles can be accelerated by differing processes depending on their energy. This picture is motivated by the vast gap that separates the kinetic scales (e.g., the gyroradius of thermal ions $r_{\rm L} = 3\times 10^3\,$cm for a $1\,$GeV proton in a magnetic field with a strength of $B=10^3\,$G) from the macroscopic scale of the source (e.g., the black hole gravitational radius $r_{\rm g}\equiv GM/c^2 \simeq 1.5\times 10^{12}\,M_7$ cm)\footnote{We use the usual shorthand notation $Q_x = Q/10^x$, where $Q$ is expressed in cgs units, with the exception of solar masses for $M$.}. In these environments, preacceleration in reconnecting current sheets is generic, and it can energize ions up to a Lorentz factor of about $\sigma$, where the magnetization parameter is defined as the ratio of the magnetic and plasma energy density~\citep{2018MNRAS.473.4840W}. Unless $\sigma$ takes very high values~\citep[e.g.,][]{2024ApJ...961L..14F,2024PhRvD.109j1306M}, particles must draw most of their energy from their interaction with the turbulence. However, as they gain energy, their mean free path increases so much that they may eventually become sensitive to the coherent shear structure of the velocity flow around the black hole and are therefore further energized through shear acceleration~\citep[e.g.,][]{2019Galax...7...78R}. 

We thus adopt this point of view and focus on the following key questions: The conditions under which these sources can push protons to $\sim30-300\,$TeV energies, and the extent to which can they account for the spectrum inferred from IceCube neutrino observations. Correspondingly, we do not seek to reproduce the 
multiwavelength spectrum observed from the prototype NGC~1068, which has been discussed and modeled to fine 
levels of detail~\citep{2020PhRvL.125a1101M,2020ApJ...891L..33I,2021ApJ...922...45K,2022ApJ...941L..17M,2022ApJ...939...43E,2023ApJ...956....8F,2024ApJ...961L..14F,2024ApJ...961L..34M,2024PhRvD.109j1306M,2024ApJ...974...75F,2024ApJ...972...44D}. However, we properly implement the associated rates of energy loss that limit the acceleration of multi-TeV to PeV protons.  

This paper is organized as follows. In Sect.~\ref{sec:acc} we discuss the implementation of particle acceleration, first in a turbulent hot corona, and then, in a sheared velocity flow. We show that various parameters control the ion energy spectra. We discuss the results in Sect.~\ref{sec:disc} and provide three scenarios that can reproduce the inferred high-energy proton spectra. One scenario considers stochastic acceleration in turbulence, another scenario combines stochastic acceleration with shear acceleration, and the third scenario involves stochastic acceleration, properly accounting for the feedback of accelerated particles on the turbulence. We emphasize that in the first two scenarios, different sources are expected to exhibit (possibly largely) different 
spectra because of the strong dependence of the derived spectra on the physical conditions (e.g., the Alfvénic velocity $v_{\rm A}$). We also argue that self-regulation of the stochastic acceleration by turbulence damping provides a well-motivated satisfactory fit to the inferred spectrum without requiring ad hoc normalization of the flux of high-energy ions. This opens new avenues of research into the physics of these cosmic accelerators. We summarize our findings in Sect.~\ref{sec:concl}.

\section{Modeling particle acceleration}\label{sec:acc}

\subsection{Radiation fields and general setup}\label{sec:setup}
We considered a generic setup in which a supermassive black hole with a mass of $M_{\rm BH}$
and size $r_g$ 
is embedded in a luminous accretion disk ($L_{\rm disk}$) corona environment. 
For simplicity, the corona that encompasses the main disk was assumed to be 
quasi-spherical and compact, with a characteristic length of $r_{\rm co} \leq 100 r_g$ 
\citep[e.g.,][]{2015MNRAS.451.4375F}. In contrast to the disk, the corona was taken to 
be hot, with proton temperatures up to virial, that is, $T_p \simeq 1.2 \times 
10^{11}~(30/\hr)$ K, where $\hr \equiv r/r_g$. 
While still high, radiative cooling results in electron temperatures that are 
significantly lower ($T_e \ll T_p$), as evidenced by the observed X-ray spectral 
cutoffs ($\epsilon_{cX}$) at tens to hundreds of keV \citep[e.g.,][]{2018ApJ...866..124K,
2022ApJ...927...42K,2024FrASS..1008056K}. In principle, energy dissipation through reconnection in 
magnetically dominated regions could represent an important means to maintain these 
electron temperatures \citep[e.g.,][]{1997MNRAS.291L..23D,2001MNRAS.321..549M,2002ApJ...572L.173L,
2020ApJ...899...52S}. 
Characteristic coronal plasma $\beta$ parameters ($\beta_p = p_{\rm gas}/p_{\rm mag}$) are 
expected to be about unity, but they likely decrease with increasing distance 
from the midplane \citep[e.g.,][]{2003ApJ...599.1238D,2005ApJ...620..878D}. 
While AGN corona have commonly been thought to be optically thin ($\tau_T< 1$; \citealt{1985ApJ...289..514Z,1995ApJ...449L..13S}, recent modeling suggested that in 
some sources, $\tau_T$ may reach values up to a few \cite[e.g.,][]{2017MNRAS.468.3489K,
2022ApJ...927...42K}.
Characteristic Alfv\'en speeds are about $v_A \simeq 0.15\, (30/\hr)^{1/2} 
\beta_p^{-1/2} c$, which makes efficient stochastic acceleration feasible\footnote{$v_{\rm A}$ sets 
the characteristic velocity of turbulent eddies on the outer scale because we assumed a strongly 
turbulent corona with $\delta B\sim B$ and $\beta_p\sim 1$.}.
In terms of $\beta_p$, the thermal plasma $\beta$ parameter, the corona magnetic field may 
reach strengths of $B=\sqrt{8\pi n_p m_p c^2/(3\hr \beta_p)} \simeq 4 \times 
10^3 ~(30/\hr) M_7^{-1/2} \beta_p^{-1/2} \tau_T^{1/2}$ G, where $\tau_T = 
n_p \sigma_T r_c$ \cite[cf.,][]{Padovani2024}. In general, however, there are 
uncertainties because the physical origin of the corona and its emission are not well understood. 
In particular, magnetic field estimates based on the detection of coronal radio synchrotron 
emission (hybrid corona), for example, rather suggest $B \sim$ O(10~G) on scales of $\sim 
80 r_g$ \citep{2018ApJ...869..114I, 2020ApJ...891L..33I}. It seems conceivable that the magnetic 
field decays away from the midplane. In the following, we take $B=10^3$ G as the reference value for 
the corona in application for NGC~1068. We discuss implications for varying fields, but we already 
note that most of our results are insensitive to this choice. Unless the magnetic field takes values 
well in excess of $10^4\,$G, proton synchrotron losses can indeed be safely neglected. 
Furthermore, the rate of acceleration in stochastic acceleration is quantified by $v_{\rm A}$ 
and $\ell_{\rm c}$ (the outer scale of the turbulence), not by $B$ per se. The strength of the 
magnetic field enters the expression for the particle mean free path, which controls escape 
losses and the efficiency of shear acceleration, but with an exponent of about one-third
(see Appendix~\ref{sec:app}).

As mentioned above, we focused on the suprathermal proton spectra produced by turbulent and/or 
shear acceleration and discuss the extent to which this can reproduce a spectrum as inferred from 
the IceCube observations of NGC~1068 \citep{2022Sci...378..538I}. This proton spectrum has a 
generic spectral index $s\sim -3$ in the energy interval $\sim 30-300\,$TeV, and its 
normalization is such that the suprathermal particle pressure at $\sim 30\,$TeV is expected to carry 
a fraction from $\sim 0.01$ to $\sim 0.3$ of the total plasma 
pressure ~\citep{2020PhRvL.125a1101M,2021ApJ...922...45K,2022ApJ...939...43E,2024ApJ...972...44D}. 
In the figures that follow, we express the energy spectra in units of $p_{\rm gas}$ and 
accordingly normalize the butterfly diagram characterizing the inferred proton energy
spectrum to $0.3$, corresponding to a pressure ratio of $0.1$. Because of the experimental and 
modeling uncertainties, this normalization is itself uncertain to within a factor of a few.

\subsection{Particle acceleration in turbulence}\label{sec:turb}
Stochastic particle acceleration has received increased attention in recent years, with 
notable inputs from numerical simulations~\citep{17Zhdankin,2017PhRvL.119d5101I,2018JPlPh..84f7201P,18Comisso,2019ApJ...886..122C,2019MNRAS.485..163K,2019PhRvL.122e5101Z,2020ApJ...893L...7W,2020ApJ...894..136T,Bresci+22,2022ApJ...928...25P,2022ApJ...936L..27C,2023ApJ...959...28P} and theoretical developments~\citep{14Lynn,2016MNRAS.458.2584B,2019PhRvD..99h3006L,2020MNRAS.499.4972L,2020PhRvD.102b3003D,2020MNRAS.491.3860S,2020ApJ...895L..14S,PhysRevD.104.063020,2022PhRvL.129u5101L,2023ApJ...942...21X}.

In phenomenological applications, stochastic acceleration is commonly modeled through 
a purely diffusive Fokker-Planck equation, which is characterized by a (energy-dependent) diffusion 
coefficient $D_{\epsilon\epsilon}$, an approach mostly motivated by simplicity and by arguments 
following a quasilinear description of weak turbulence theory \citep[e.g.,][]{1989ApJ...336..243S}. In this 
approach, all information regarding the acceleration rate is encoded in the diffusion coefficient $D_{\epsilon\epsilon}$, 
in particular, the acceleration rate $\nu_{\rm acc}\equiv D_{\epsilon\epsilon}/\epsilon^2$. The general transport equation that 
describes the evolution of the particle distribution $n_\epsilon\equiv {\rm d}n/{\rm d}\epsilon$ then takes the form
\begin{equation}
    \partial_t n_\epsilon\,=\,\mathcal L_{\rm stoch}n_\epsilon - \partial_\epsilon\left(\dot \epsilon_{\rm loss} n_\epsilon\right) 
    - \frac{n_\epsilon}{\tau_{\rm esc}}\,,
    \label{eq:transport}
\end{equation}
where $\mathcal L_{\rm stoch}$ denotes the diffusive operator describing stochastic 
acceleration,  that is, 
\begin{equation}
    \mathcal L_{\rm stoch} n_\epsilon \equiv \partial_\epsilon\left(D_{\epsilon\epsilon}\partial_\epsilon n_\epsilon\right) - 
2\partial_\epsilon\left(D_{\epsilon\epsilon} n_\epsilon/\epsilon\right)
\label{eq:FP-stoch}
\end{equation} 
for the purely diffusive Fokker-Planck scheme. In Eq.~(\ref{eq:transport}), $\dot \epsilon_{\rm loss}$ 
denotes the proton energy loss rate (here, a negative quantity by convention) associated with Bethe-Heitler 
pair production $p\gamma\rightarrow e^-e^+$, and hadronic $p$-$p$ and $p$-$\gamma$ interactions that 
lead to neutrino production. These energy losses become significant at energies $\epsilon\gtrsim 10\,$TeV (
see, e.g.,~\cite{2022ApJ...941L..17M}), and their implementation is discussed in Appendix~\ref{sec:app}. 

The timescale $\tau_{\rm esc}$ characterizes the time over which particles escape from the acceleration region through 
spatial diffusion. In a highly dynamic turbulence, it is important to recall that particle transport can occur 
both through scattering on magnetic inhomogeneities (with mean free path $\lambda_{\rm scatt}$) and through turbulent 
transport, with a characteristic diffusion coefficient $\kappa_{\rm turb}\sim \ell_{\rm c}v_{\rm A}/3$. As the scattering 
mean free path is about $\lambda_{\rm scatt}\sim r_{\rm L}^{1/3}\ell_{\rm c}^{2/3}$ (see Appendix~\ref{sec:app}), 
diffusive escape is in practice governed by turbulent transport. This point seems to have gone mostly unnoticed, even 
though it brings an important constraint on the geometry of the corona. A large $\ell_{\rm c}$ diminishes the acceleration 
rate, but increases the escape rate, while a high $v_{\rm A}$ increases both acceleration and escape rates. For 
clarity, we adopted a diffusion coefficient $\kappa = \kappa_{\rm turb} + \lambda_{\rm scatt}c/3$ and discuss 
the consequences in Sect.~\ref{sec:disc}. The diffusive escape timescale is then written $\tau_{\rm esc}\equiv r_{\rm co}^2/(2\kappa)$. 

Additionally, we also took the influence of particle escape from the accelerating region 
through advective transport into account. To do this, we bound the integration time to $\tau_{\rm adv}$, which is defined 
in terms of the advection velocity $v_{\rm adv}$ as $\tau_{\rm adv}\equiv r_{\rm co}/v_{\rm adv}$.
Thus, we did not seek from the outset a stationary solution to the transport equation, Eq.~(\ref{eq:transport}). 
In our formulation, the proton spectrum achieves stationarity because the injection is itself stationary in 
time. Specifically, at any given time $t$, the proton spectrum contained in the turbulent corona consists of 
particles that have been injected at times $t'\in[t-\tau_{\rm adv},\,t]$ and accelerated for a duration 
$\tau \equiv t-t'$. This spectrum can be expressed as
\begin{equation}
    \frac{{\rm d}n_{\rm cor}}{{\rm d}\epsilon}\,=\,\int_{0}^{\tau_{\rm adv}}{\rm d}\tau\,\int_0^{+\infty}{\rm d}\epsilon'\,G_{\rm turb}\left(\epsilon;\,\epsilon',\,\tau\right) \frac{{\rm d}\dot n_{\rm inj}}{{\rm d}\epsilon'}\,
    \label{eq:green-stoch}
\end{equation}
in terms of the Green function $G_{\rm turb}\left(\epsilon;\,\epsilon',\,\tau\right)$ , which is associated with Eq.~(\ref{eq:transport}). This determines the probability of a transition from $\epsilon'$ to $\epsilon$ over a time interval $\tau$, and of the injection 
rate spectrum ${\rm d}\dot n_{\rm inj}/{\rm d}\epsilon$. We assumed that the latter is constant in time, and as discussed below, that it is monoenergetic without loss of generality. 

The above implementation accommodates injection of particles at one side of the corona, then advection, or uniform 
stationary injection throughout the corona equally well. In both cases, ${\rm d}n_{\rm cor}/{\rm d}\epsilon$ can be regarded as the stationary 
average spectrum. In the former case, the time-dependent spectrum at location $x$ can be written as ${\rm d}n_{\rm turb}(x)/{\rm d}\epsilon 
= \int{\rm d}\epsilon'\,G_{\rm turb}(\epsilon;\,\epsilon',\,x/v_{\rm adv}) {\rm d}n_{\rm inj}/{\rm d}\epsilon'$ in terms of the 
injection density ${\rm d}n_{\rm inj}/{\rm d}\epsilon'$ at the initial boundary, and the substitutions $x\rightarrow v_{\rm adv}\tau$, 
${\rm d}n_{\rm inj}/{\rm d}\epsilon' \rightarrow \tau_{\rm adv}{\rm d}\dot n_{\rm inj}/{\rm d}\epsilon'$ 
lead to Eq.~(\ref{eq:green-stoch}) above, where the time integral represents an average over the corona. In the latter case, 
particles are injected throughout the corona at a rate ${\rm d}\dot n_{\rm inj}/{\rm d}\epsilon'$ (independent of $x$), so that
\begin{equation}
    \frac{{\rm d}n_{\rm turb}(x)}{{\rm d}\epsilon}\,=\,\int_0^{x}\frac{{\rm d}x'}{v_{\rm adv}}\int_0^{+\infty}{\rm d}\epsilon'\,G_{\rm turb}(\epsilon;\,\epsilon',\,(x-x')/v_{\rm adv})\,\frac{{\rm d}\dot n_{\rm inj}}{{\rm d}\epsilon'}\,.
\end{equation}
Then, ${\rm d}n_{\rm cor}/{\rm d}\epsilon$ is again recovered by taking the average of ${\rm d}n_{\rm turb}(x)/{\rm d}\epsilon$ 
over $x$ throughout the corona. 

The advection term is sometimes included in the transport equation as an escape term of the form $-n_\epsilon/\tau_{\rm adv}$, 
which implicitly implies that particles can spend a time $>\tau_{\rm adv}$ in the acceleration region and thus become accelerated 
to much higher energies. However, advection is a systematic and not a random loss, and it must therefore be treated differently. 
This impacts the particle spectra in a significant way. In the present description, in particular, advection effectively cuts 
off the spectrum at a maximum energy set by $\nu_{\rm acc}\tau_{\rm adv}\sim 1$ (or lower, if other losses dominate).

Recent numerical experiments of stochastic particle acceleration, which either tracked test particles in a magnetohydrodynamic (MHD) simulation 
or relied on a full kinetic description using the particle-in-cell (PIC) technique, have revealed a richer and more complex 
landscape than the simple Fokker-Planck formulation above. These simulations indicate a diffusion coefficient of the form
$D_{\epsilon\epsilon}\simeq 0.2  \gamma^2 (v_{\rm A}/c)^2 c/\ell_{\rm c}$ for particles with a Lorentz factor $\gamma\equiv 
\epsilon/mc^2 $, where $v_{\rm A}$ denotes the Alfvén velocity in the total magnetic field, and $\ell_{\rm c}$ is the coherence 
length of the turbulence. However, they also demonstrated that the purely diffusive Fokker-Planck approach does not successfully 
account for the particle spectra~\citep{2017PhRvL.119d5101I,2018JPlPh..84f7201P,2020MNRAS.499.4972L}, unless an 
energy-dependent (as well as $v_{\rm A}$-dependent) advection coefficient is added, which must be extracted from numerical 
simulations~\citep{2020ApJ...893L...7W}. This is not a trivial issue because the results 
of these kinetic simulations must be extrapolated far in time in order to make the connection with phenomenology. For instance, for our fiducial values 
$v_{\rm adv}=0.02\,c$, $r_{\rm co}=30\,r_{\rm g}$ and $\ell_{\rm c}=10\,r_{\rm g}$, the advection timescale 
reads $\tau_{\rm adv}\simeq 150\ell_{\rm c}/c$, while PIC simulations typically run over $\simeq 10\,\ell_{\rm c}/c$.

In \citet{PhysRevD.104.063020,2022PhRvL.129u5101L}, we argued that particles are accelerated through a generalized Fermi 
process, in which particles gain energy as they cross intermittent regions of dynamic, curved, and/or compressed magnetic field 
lines. This model was benchmarked on MHD simulations in a regime relevant to black hole coronae, namely a sub- to mildly 
relativistic Alfvénic velocity, and the time-dependent Green functions for the spectra obtained in this approach also match 
those seen in PIC simulations in similar conditions~\citep{2022ApJ...936L..27C}. To account for the intermittent nature of the 
accelerating structures, it proves necessary to go beyond a purely diffusive Fokker-Planck treatment and to consider the full 
probability distribution function of the random forces acting on the particles. The transport can then be modeled using 
Eq.~(\ref{eq:transport}) above, with a more general diffusion operator characterized by 
\begin{equation}
    \mathcal L_{\rm stoch}n_\epsilon\,\equiv\,\int_{0}^{+\infty}{\rm d}\epsilon'\,
\left[\frac{\varphi\left(\epsilon\vert \epsilon'\right)}{t_{\epsilon'}} n_{\epsilon'}(t) - 
\frac{\varphi\left(\epsilon'\vert \epsilon\right)}{t_{\epsilon}} n_\epsilon(t)\right]\,,
\label{eq:stoch_CK}
\end{equation}
where $\varphi\left(\epsilon\vert \epsilon'\right)$ represents the kernel describing the probability of jumping 
from $\epsilon'$ to $\epsilon$ over a time interval $t_{\epsilon'}$, as described in \citet{2022PhRvL.129u5101L} 
(see also \citet{2017PhRvL.119d5101I} for related considerations). This kernel was determined through dedicated 
measurements carried out in an MHD simulation with an Alfvénic velocity $v_{\rm A}=0.4\,c$. It can nevertheless be 
extrapolated to other Alfvénic velocities by rescaling its width in proportion to $(v_{\rm A}/0.4\,c)^2$, and we do 
this below. 

In order to leverage these recent findings and to gauge the influence of the underlying theoretical 
uncertainty on the predicted particle spectra, we adopt both pictures in the following, that is, we 
integrate the transport equation, Eq. (\ref{eq:transport}), using either the Fokker-Planck kernel 
[Eq.~(\ref{eq:FP-stoch}), hereafter referred to as model (1)] or that describing the generalized Fermi 
process [Eq.~(\ref{eq:stoch_CK}), hereafter model (2)]. Our main motivation here is to provide a more 
accurate description of the proton and secondary neutrino spectra and of the corresponding acceleration rate.

For practical applications, the generalized Fermi process produces time-dependent 
Green functions that take the form of broken power laws in the absence of escape and losses. 
These solutions obviously differ from the lognormal shape \citep[e.g.,][]{1962SvA.....6..317K} 
that follows from a purely diffusive Fokker-Planck kernel with $D_{\epsilon\epsilon}\propto 
\epsilon^2$ (in the absence of escape and losses). When a continuous injection of particles is 
included, both formulations nevertheless lead to hard spectra beyond the injection Lorentz factor 
$\gamma_0$~\citep{1984A&A...136..227S}, up to a maximum energy that is either dictated by the 
finite time ($\tau_{\rm adv}$) over which particle acceleration takes place, or by energy losses. 

We do not discuss the physics of injection here and simply assumed that a given number of particles was 
injected into turbulent acceleration with a characteristic Lorentz factor $\gamma_0\sim 1$. The value of 
this Lorentz factor, or the exact shape of the injected spectrum, is not relevant because turbulent acceleration 
effectively loses the dependence on the initial conditions when an acceleration to high energies has taken place. 
This injection could either take place through magnetic reconnection in localized patches of sufficient 
magnetization~\citep[e.g.,][]{2024PhRvD.109j1306M} through fast, small-scale turbulent acceleration in 
localized regions, as observed in some simulations~\citep[e.g.,][]{2018JPlPh..84f7201P,2020ApJ...894..136T}, 
or simply because a fraction of the energy density is contained in supra-thermal particles accelerated 
elsewhere and accreted into the corona. The exact energy fraction contained in these particles is relevant, 
however, as we discuss in the following.

\begin{figure}[t!]
\centering
\includegraphics[width=0.45\textwidth]{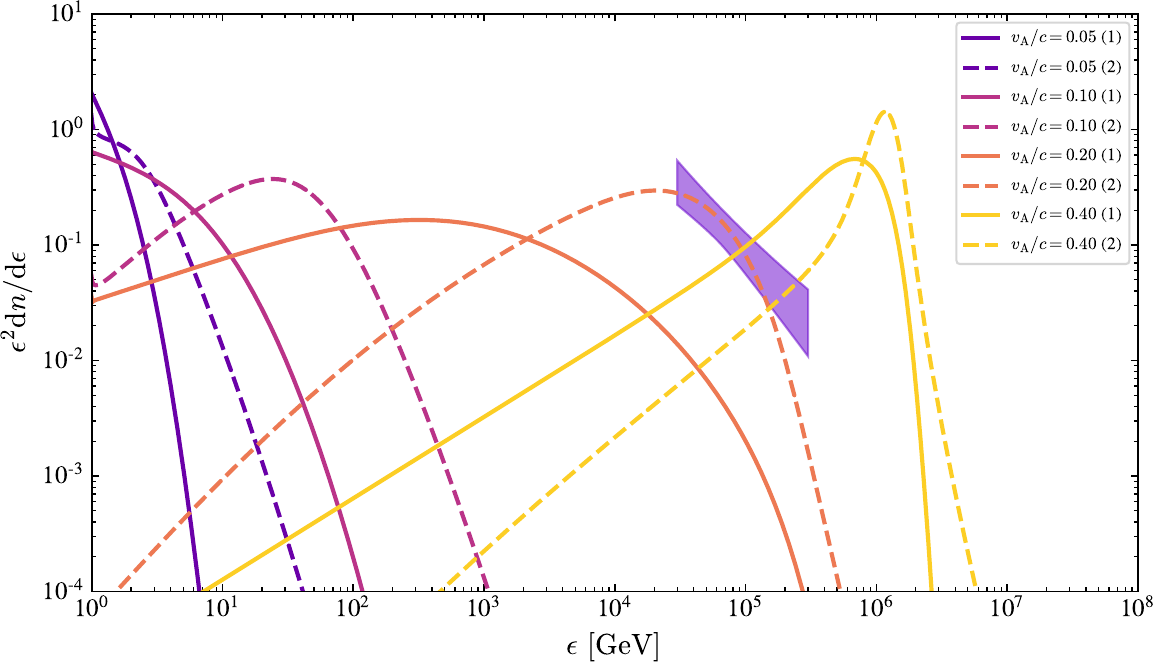}
\includegraphics[width=0.45\textwidth]{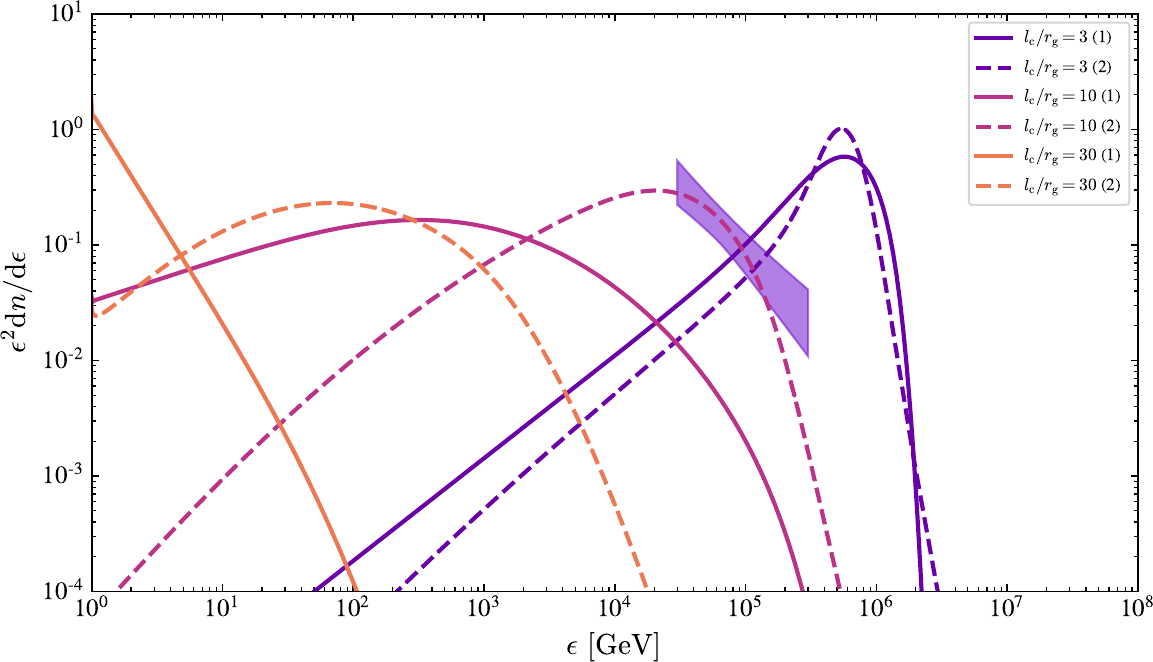}
\includegraphics[width=0.45\textwidth]{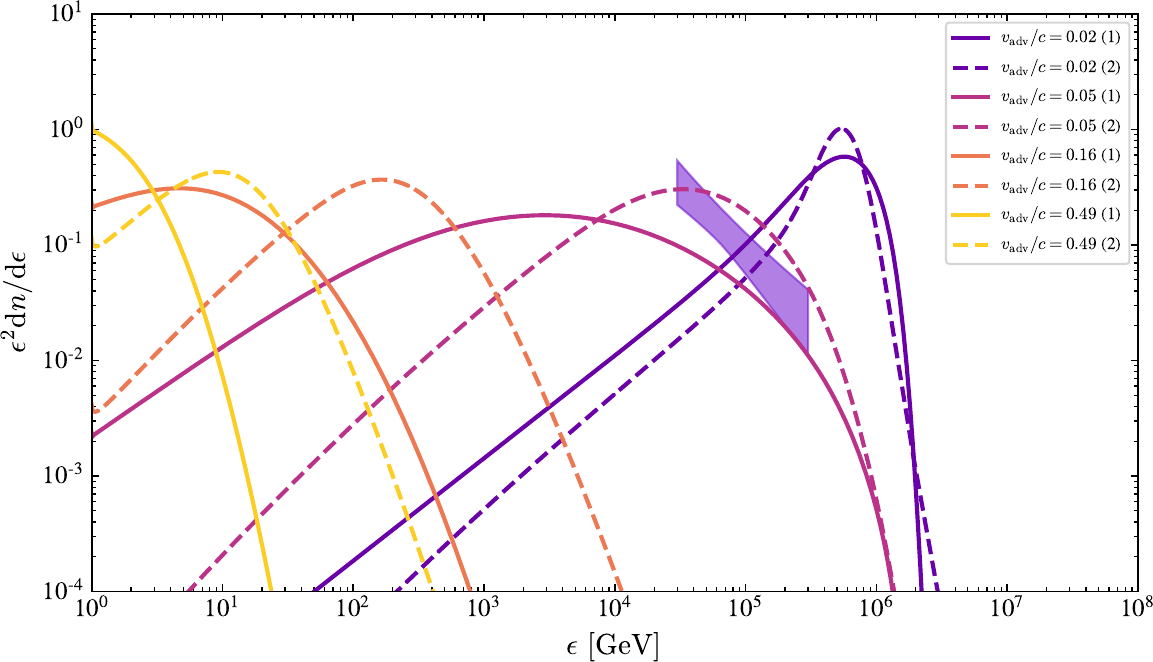}
\caption{Proton energy spectra $\epsilon^2\,{\rm d}n_{\rm cor}/{\rm d}\epsilon$ (per log-interval of energy) vs. 
proton energy $\epsilon$ predicted by stochastic acceleration in a turbulent corona, starting from mono-energetic 
protons with a Lorentz factor of $\gamma_0\sim1$. In each panel, the solid line corresponds to a solution obtained by 
integrating the Fokker-Planck equation up to time $\tau_{\rm adv}=r_{\rm co}/v_{\rm adv}$ [model (1) including energy 
losses, advection, and escape as described in the text], and the dashed line shows the corresponding solution for 
the same parameters for the generalized Fermi model described by Eq.~(\ref{eq:stoch_CK}) [model (2)]. In all panels, 
$r_{\rm co}$ is set to $30r_{\rm g}$, and in the top and middle panels, $v_{\rm adv}\simeq 0.02\,$c. The upper panel shows 
the dependence on $v_{\rm A}$, assuming otherwise a coherence length scale for the turbulence $\ell_{\rm c}=
10\,r_{\rm g}$ and an advection velocity $v_{\rm adv}=0.02\,c$. From left to right, as indicated, $v_{\rm A}/c=0.05$, 
$v_{\rm A}/c=0.10$, $v_{\rm A}/c=0.20$, and $v_{\rm A}/c=0.40$. The middle panel examines the dependence on $\ell_{\rm c}$, 
assuming otherwise $v_{\rm A}/c=0.20$ and $v_{\rm adv}/c=0.02$. From right to left, as indicated, $\ell_{\rm c} = 
3\,r_{\rm g}$, $\ell_{\rm c} = 10\,r_{\rm g}$, and $\ell_{\rm c} = 30\,r_{\rm g}$. The bottom panel examines the influence 
of advection, characterized here by the advection velocity $v_{\rm adv}$ (which sets the advection timescale $\tau_{\rm adv}
\equiv r_{\rm co}/v_{\rm adv}$), assuming otherwise $v_{\rm A}/c=0.20$ and  $\ell_{\rm c} = 3\,r_{\rm g}$. From right to 
left, as indicated, $v_{\rm adv}/c=0.02$, $v_{\rm adv}/c=0.05$, $v_{\rm adv}/c=0.16$ and a (more extreme) case 
$v_{\rm adv}/c=0.49$. The butterfly indicates the range of values needed to reproduce a neutrino spectrum as observed 
by IceCube for NGC~1068. The energy density spectrum is expressed in units of the background plasma pressure (see 
Sect.~\ref{sec:setup}). In these units, each individual spectrum has been normalized to an integrated fraction of unity 
for the sake of illustration.
\label{fig:spec_stoch1}}
\end{figure}

The spectral shape is controlled by the following parameters: $D_{\epsilon\epsilon}$ or 
$\varphi\left(\epsilon\vert
\epsilon'\right)$ for the acceleration rate. They both depend on $v_{\rm A}$, $\epsilon$ and
$\ell_{\rm c}$; $\tau_{\rm adv}\equiv r_{\rm co}/v_{\rm adv}$ , which sets the integration time before advective 
escape at velocity $v_{\rm adv}$, removes the particles $\tau_{\rm esc}$ and $\tau_{\rm loss}$, which control the
timescales associated with diffusive escape and radiative energy losses. We followed previous 
treatments for the latter and recall their value in Appendix~\ref{sec:app}. Figure~\ref{fig:spec_stoch1} shows the results 
obtained for the Fokker-Planck model (solid lines) and the generalized Fermi process (dashed lines) for various 
choices of $v_{\rm A}$ (upper panel), $\ell_{\rm c}$ (middle panel), and $v_{\rm adv}$ (lower panel). The butterfly 
diagram shows the range of values inferred from the IceCube detection of neutrinos from NGC~1068, as discussed above. The energy spectra shown here is expressed in units of the background plasma 
pressure (Sect.~\ref{sec:setup}). However (see also below), the spectra shown here were normalized in an 
ad hoc way to a total energy density of unity in these units. In general, turbulent acceleration allows for a 
favorable situation by yielding sufficiently hard spectra, which effectively minimizes the required energy 
budget if the peak occurs close to the observed energy range.

The trivial observation that hardly any of the spectra fit this 
butterfly diagram can be misleading. 
The important point is that the predicted spectra are extremely sensitive to the choice of parameters. 
This is easily understood, noting that for $D_{\epsilon\epsilon}\propto \epsilon^2$, the mean energy 
increases exponentially fast in time, that is, $\langle \epsilon\rangle \propto \exp\left(4\nu_{\rm acc}
t\right)$, with $\nu_{\rm acc}\equiv D_{\epsilon\epsilon}/\epsilon^2$ the acceleration rate defined earlier. 
Consequently, in a fixed time interval $t=\tau_{\rm adv}$ over which $\nu_{\rm acc} t\gg1$ (needed to 
achieve acceleration to high energies), changing $\nu_{\rm acc}$ by a factor of a few changes the maximum 
energy by orders of magnitude. The dependence of $\nu_{\rm acc}$ on the parameters is recalled above, 
$\nu_{\rm acc} \propto v_{\rm A}^2/ \ell_{\rm c}$. We assumed $\delta B/B\gtrsim 1$, otherwise  
$\nu_{\rm acc}$ should be further suppressed by a factor $(\delta B/B)^2$. The dependence on the advection 
velocity is also easily understood because the number of $e-$folds of the increase in the maximum energy 
scales as $\nu_{\rm acc} \tau_{\rm adv}\propto 1/v_{\rm adv}$. From this, we can also infer the dependence 
on the scale of the corona. The spectra that reach PeV energies are cut off because of catastrophic hadronic
losses. Other spectra are cut off because of the limited time they spend in the corona. 

Clearly, parameter values that fit the inferred butterfly diagram might be found, and examples are 
shown below. This comes at the price of some fine-tuning, however. One important corollary is that 
under the above assumptions, we do not expect all AGNs, even those with a similar X-ray luminosity, 
to produce neutrinos with a similar spectrum. 
If the physical conditions that dictate the rate of acceleration (e.g., $v_{\rm A}$) show an 
extended distribution over the neutrino-emitting AGN subpopulation, the population as a whole is instead expected to 
produce an extended neutrino energy spectrum, from $\lesssim\,$GeV to $\sim 0.1\,$PeV. This has
consequences for our prediction of the diffuse flux from this population of AGNs and for the modeling 
of the spectrum of a given source, as we discuss further in Sect.~\ref{sec:disc}.

Another important lesson drawn from Fig.~\ref{fig:spec_stoch1} is that fitting the inferred proton 
spectrum raises an energy budget issue, as noted earlier (Sect.~\ref{sec:setup}), which also represents an 
intriguing coincidence. Namely, in order to match the flux, we must inject 
a tiny fraction of the thermal plasma into stochastic acceleration, such that after energization by a factor $\sim 10^5$, its pressure is comparable to that of its parent population. If more particles had been injected, the 
energy requirement would have become excessive, while in the opposite case, the proton flux falls short 
of the required flux. Moreover, if suprathermal particles carry an energy density comparable to that in the 
thermal component, they must backreact on the flow, and in particular, damp the turbulence that feeds them 
in energy. In this situation, stochastic acceleration 
becomes nonlinear and self-regulated by damping~\citep{1979ApJ...230..373E,1979ApJ...229..413E,
2012A&A...544A..94L,2016ApJ...816...24K,2024PhRvD.109f3006L}.  We discuss this possibility further 
in Sect.~\ref{sec:disc}.

Alternatively, the turbulent corona might not shape the entire proton spectrum, but only part 
of it, and protons escaping the corona might be further accelerated in the sheared velocity flows at the 
base of the jet. We first discuss how this shear acceleration can be implemented, and we then examine the general
combination of the two scenarios in Sect.~\ref{sec:disc}.

\subsection{Particle acceleration in sheared flows}\label{sec:shear}
Accreting black hole systems are known to drive relativistic jets and winds, and thereby provide an environment 
that is conducive to shear-type particle acceleration \citep[e.g.,][]{2018PhRvD..97b3026K,2019Galax...7...78R,2019PhRvD..99h3006L,
2022ApJ...933..149R,2023ApJ...958..169W,2024ApJ...967...36W}.
This might provide a complementary mechanism for proton energization if 
turbulent acceleration in the corona were to saturate at multi-TeV energies.
To be efficient, shear acceleration commonly requires seed injection of energetic particles, 
which is in line with a hierarchical acceleration scenario. In our context, we envisaged that 
the black hole vicinity supports a fast, sheared outflow; for example, that the hot X-ray emitting 
corona forms the base of an outward-moving jet, which may be supported by radiation pressure or be driven 
magnetically \citep[e.g.,][]{1999MNRAS.305..181B, 2001MNRAS.326..417M,2002MNRAS.332..165M,
2017ApJ...835..226K,2022ApJ...935L...1L,2023ApJ...949L..10P,2024MNRAS.528L.157D,2025ApJ...979..199S}. 
For the prototype Seyfert galaxy NGC~1068, jet-like features are indeed apparent on larger 
scales, although jet speeds are most likely nonrelativistic ($\lesssim 0.1$c) on scales of 
several dozen parsec ($\sim 10^7 r_g$) from the nucleus \citep{1996ApJ...458..136G,
2000evn..conf....7R,2017MNRAS.469..994M}. Part of its observed gamma-ray emission may in 
fact be related to the jet \citep[e.g.,][]{2010A&A...524A..72L, 2024A&A...687A.139S}. 
The radio-inferred (large-scale and time-averaged) jet power is about 
$L_{\rm jet} \sim 10^{43}$ erg/s \citep{Padovani2024}. On the other hand, the 
minimum (isotropic-equivalent) cosmic-ray 
power required to account for the observed neutrino emission is about several 
$10^{42}$ erg/s \citep[e.g.,][see also Appendix~\ref{sec:app}]{2022ApJ...941L..17M}. If 
part of this emission is indeed related to the outflow, even if it were relativistically boosted, it 
seems conceivable that cosmic-ray acceleration may contribute to the deceleration of the  flow 
(viscous drag) as it propagates outward. We did not include this 
possible backreaction and self-regulation of the acceleration process in our modeling, but we note that it 
may impact on the bulk flow profile (e.g., shear broadening) and on the turbulence generation. 

\begin{figure}[t!]
   \centering
   \includegraphics[width=0.5\textwidth]{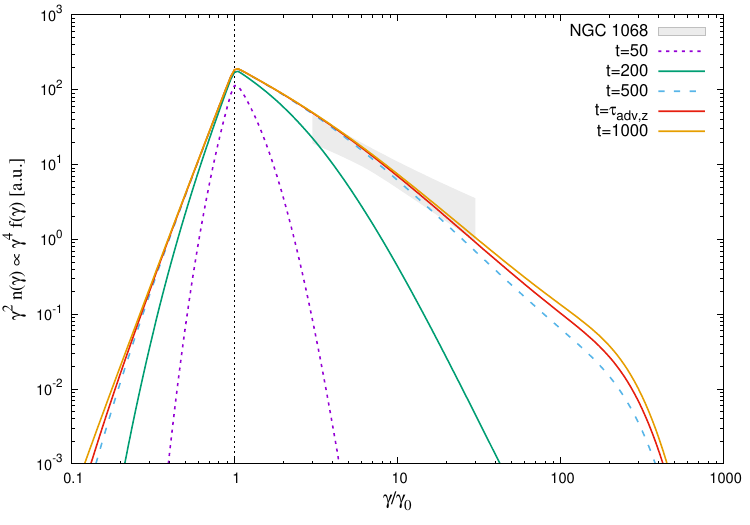}
   \caption{Exemplary time-dependent proton energy distribution obtained from jet-shear acceleration 
   with a Fokker-Planck model assuming continuous injection of seed protons with $\gamma_0 =10^4$.
   As a result of advective escape, the predicted particle spectrum will appear somewhat softer 
   (red, $t=\tau_{\rm adv,z}$) compared to the otherwise quasi steady-state distribution (yellow).
   The gray shaded region illustrates the spectral characteristics required to reproduce the neutrino 
   emission for NGC~1068 as measured by IceCube.}
   \label{fig:acc_shear}
\end{figure}

For reference, we explore a model below in which turbulent acceleration in the corona 
saturates at $\gamma_{p} \leq 10^4$ (10 TeV) either due to $v_A \lesssim 0.1\,c$  or $\ell_{\rm c} \gg 
r_g$ in the corona. The corona thus sustains a reservoir of energetic seed particles that 
can diffuse into the outflow and experience further jet-shear acceleration. In this process, 
acceleration is related to a particle that effectively samples the velocity difference while it is 
scattered across a shearing flow \citep{2019Galax...7...78R}. For simplicity, we considered a 
collimated mildly relativistic outflow or jet, with a Lorentz factor of $\gamma_{j} \equiv 1/(1-u_{z}^2/c^2)^{1/2} 
\lesssim 3$, characterized by a lateral velocity shear $u_z(r)$ with a width of $\Delta r = \hat{r}_{\rm sh} r_g$. 
Energetic protons undergoing shear acceleration are subjected to radiative 
losses and diffusive escape (see Appendix~\ref{sec:app}). For the phenomenological application, the 
acceleration process can be modeled through a diffusive Fokker-Planck equation for $n(\epsilon)$
of the type Eq.~(1), with mean momentum diffusion coefficient $D_{\epsilon\epsilon} \propto p^2 
\bar{D}_p$, where for energetic protons, $\epsilon=pc=\gamma m_p c^2$. 
The coefficient $\bar{D}_{p} \equiv a_g \lambda/c$ can be obtained from the local ($r$-dependent) 
shear coefficient $D_{p}(r)=(1/15) \gamma_j(r)^4 (\partial u_z/\partial r)^2 \lambda/c$ by 
suitable spatial averaging, for example, $a_g = (2/3) 
(c/\Delta r)^2/w$ with $w = 116\, (\ln[(1+\beta_0)/(1-\beta_0)])^{-2}$ in the case of a linearly 
decreasing shear flow profile with a maximum (spine) speed $\beta_0 = (1-1/\gamma_{j,\rm max}^2)^{1/2}$
\citep{2019ApJ...886L..26R,2022ApJ...933..149R}. 
In the former expression, $\lambda = c \tau_c = \eta\,(\Delta r)\left(r_L/\Delta r\right)^{1/3}$ 
denotes the scattering mean free path ($r_L$ the gyroradius), for which we took a 
scaling with $\eta =2$, following \cite{2019ApJ...886L..26R} and \cite{2023MNRAS.519.1872W} (see 
also Appendix~\ref{sec:app}). 
Particles can escape from the acceleration region either due to cross-field diffusion (laterally), 
that is, $\tau_{\rm esc}=(\Delta r)^2/(2\kappa)$ with $\kappa=\lambda c/3$, or through advective transport 
(along $z$), $\tau_{\rm adv,z}$. The accelerated 
particle distribution on scales of the corona is of interest here. While shear acceleration likely continues to 
operate along the jet, we considered the former to dominate hadronic interactions and thus neutrino 
production. Accordingly, we approximated advective escape by $\tau_{\rm adv,z} \simeq z/\langle u_z 
\rangle = \tilde{z} r_{\rm co}/\langle u_z \rangle$, with $\langle u_z \rangle$ the mean velocity 
(corresponding to $\beta_0 c/2$), and $\tilde{z}\equiv z/r_{\rm co}$ 
lower than a few  here to ensure that the target photon density is high enough. As before (Sect.~\ref{sec:turb}), 
this was implemented as a constraint on the time available for particle acceleration. 
Figure~\ref{fig:acc_shear} presents exemplary proton energy spectra as a function of time (in 
units of $\tau_c(\gamma_0)$) assuming that seed protons are continuously injected with a Lorentz 
factor of $\gamma_0=10^4$ into a shearing flow with $\gamma_j =3$, $B=50$ G, $\hat{r}_{\rm sh}=2$ and 
$\tilde{z}=2$, assuming a corona size of $r_{\rm co}=30 r_{\rm g}$. 

In general, the predicted spectra are sensitive to the choice of parameters, for example, the time 
available for particle acceleration (set by $\tau_{\rm adv,z}$). One important corollary of 
the latter is that shear modeling imposes constraints on the size and geometry of the corona. 
In addition, the required maximum outflow speed to model the CR parent distribution behind 
the IceCube neutrino spectrum is sensitive to the flow profile. In comparison to power-law or 
Gaussian-type velocity profiles, a linearly decreasing shear flow profile requires higher 
speeds to reproduce hard particle spectra. For the former, $\gamma_j\leq 2$ are 
sufficient \citep{2022ApJ...933..149R}. This suggests that mildly relativistic outflow 
speeds are sufficient to account for an inferred proton spectrum such as in NGC~1068. 
Jet deceleration due to cosmic-ray-induced viscous drag, interaction 
with a central magnetic or cosmic-ray-driven wind, along with the efficient generation 
of turbulence and entrainment of ambient material, may explain the relatively low velocity 
of the jet in NGC~1068 on larger scales. 
Our current model at this stage remains exploratory in nature since the innermost outflow 
characteristics in Seyfert-2 AGNs remain uncertain, and since we neglected a possible shear 
contribution due to flow rotation. The latter may facilitate transport of angular momentum 
from the underlying accretion flow, and it night in effect contribute to further spectral hardening 
\citep[][]{2002A&A...396..833R}. In general, while particle acceleration in sheared flows 
will not lead to a canonical power-law shape, faster flows will (ceteris paribus) result in 
harder particle spectra. This agrees with the above corollary, according to which 
not all X-ray bright Seyfert AGNs are expected to have similar neutrino spectra.

\section{Discussion}\label{sec:disc}
In the preceding section, we have examined the general features of particle acceleration in a 
turbulent corona, with possible reacceleration in the sheared velocity flow outside of this 
corona. We interpret these results here and combine them to suggest ways in which the generic 
proton spectrum inferred from IceCube observations can be reproduced, and we draw conclusions regarding the phenomenology. 
More specifically, we investigated the following three scenarios.

First, we considered pure stochastic acceleration in the turbulent corona, as modeled in 
Sect.~\ref{sec:turb}, and we demonstrate that a reasonable set of parameters can be found to reproduce 
the high-energy general slope. This was known from previous studies~\citep[e.g.,][]
{2020PhRvL.125a1101M,2020ApJ...891L..33I,2021ApJ...922...45K,2022ApJ...941L..17M,2022ApJ...939...43E,
2024ApJ...974...75F}; the main novelty here lies in the transport equation we used, in particular, for the implementation of generalized Fermi acceleration, and in the treatment of 
escape and advection.
As noted above (Sect.~\ref{sec:turb}), matching the spectrum requires some tuning in terms of 
the physical parameters and in terms of flux normalization. This suggests that other sources might 
display vastly different neutrino spectra even for otherwise similar coronal X-ray luminosities. 
Interestingly, the IceCube experiment reports differing spectra from other Seyfert galaxies 
\citep[e.g.,][]{2025ApJ...981..131A,2024arXiv240607601A}. In the case of NGC~4151 ($d\sim 16$ Mpc), 
for instance, the characteristic neutrino energy seems higher by a factor $\sim 5$, and the high-energy 
slope is slightly harder than in NGC~1068. 

\begin{figure}[t!]
\centering
\includegraphics[width=0.45\textwidth]{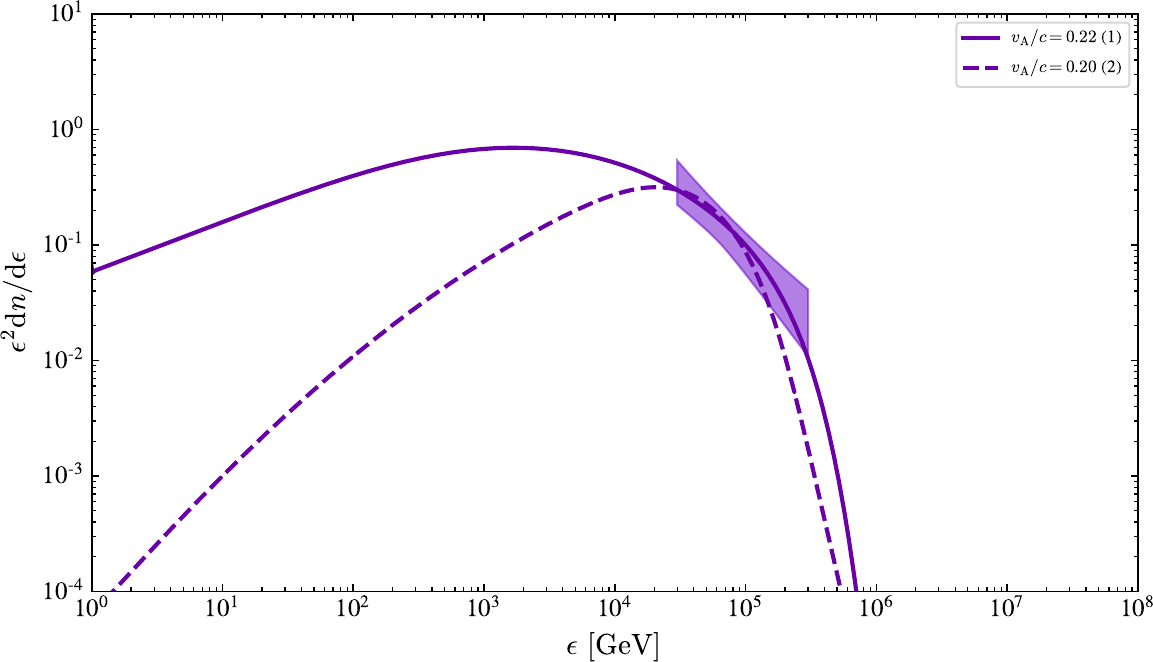}
\includegraphics[width=0.45\textwidth]{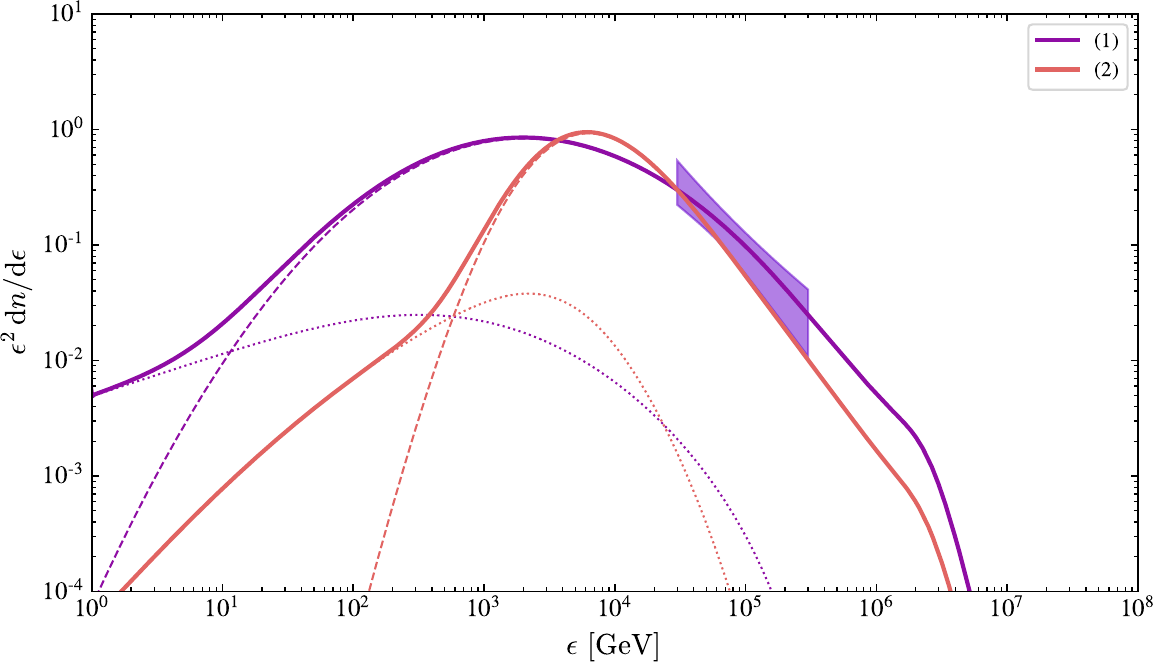}
\includegraphics[width=0.45\textwidth]{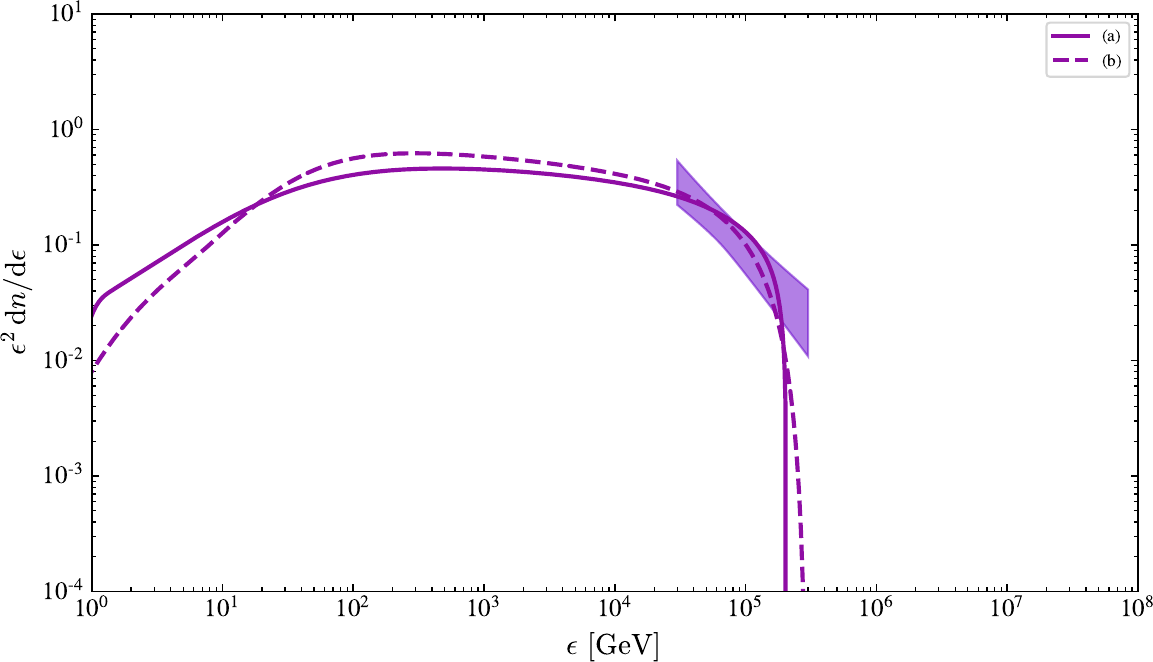}
\caption{Proton energy spectra ${\rm d}u_{\rm p}/{\rm d}\ln\epsilon\equiv\epsilon^2\,{\rm d}n/{\rm d}\epsilon$ 
vs. proton energy $\epsilon$ for three characteristic scenarios described in the text. For each panel, the 
energy density distribution is shown in units of the background plasma pressure. Top panel: Stochastic acceleration 
in the turbulent corona, with a characteristic Alfvénic velocity $v_{\rm A}\simeq 0.2\,c$ (see text for details), for 
model (1) and (2), corresponding to the solid and dashed lines, respectively. Middle panel: Combination of stochastic 
acceleration in the corona with shear acceleration at the base of the jet (see text for the details of the 
physical parameters). The stochastic contribution is indicated by dotted lines, the shear contribution 
by dashed lines, and the sum of the two by solid lines. The different colors correspond to models (1) and (2). 
For the top and middle panels, the normalization of the energy density in accelerated particles has been fixed 
in order to reproduce the inferred proton flux. Bottom panel: Modeling of stochastic acceleration, accounting 
for self-regulation by particle feedback on the turbulence (damping), for a characteristic Alfvén velocity 
$v_{\rm A}\simeq 0.2\,c$. The plasma 
$\beta$ parameter is $0.5$ (see Appendix~\ref{sec:app}).
\label{fig:combined}}
\end{figure}

The spectra obtained in this first scenario are shown in the top panel of Fig.~\ref{fig:combined}. While the parameters are the same for the two spectra shown, namely $r_{\rm co}
=30\,r_{\rm g}$, $\ell_{\rm c}=10\,r_{\rm g}$, $v_{\rm adv}=0.02\,c$ and $v_{\rm A}\simeq 0.2\,c$, their flux normalizations are different. The ratio of integrated energy density of suprathermal particles to 
background plasma pressure is $\simeq 4.3$ for model (1) and $1.0$ for model (2). Because 
the ratio of the pressures is lower by a factor $3$, the dynamical influence of the suprathermal particles should not 
be ignored in model (1). This further motivates the need to study the impact of backreaction on the turbulence, 
which is discussed further below. Finally, we note that other sets of parameters might be found, for example, a lower 
Alfvén velocity with a smaller coherence length scale. However, the general spectral shape should be 
preserved.

In a second scenario, we investigated the interplay of turbulent and shear acceleration. Shear acceleration 
requires high-energy seed particles because the rate at which acceleration proceeds increases with the particle 
mean free path~\citep[e.g.][]{2019Galax...7...78R,2022ApJ...933..149R,2023ApJ...958..169W,2024ApJ...967...36W}, 
so that in the absence of turbulent preacceleration, this mechanism would be effectively inefficient. In our setting, particles that have been preaccelerated in a turbulent corona are assumed to enter a region 
of sheared velocity flow, either through diffusive escape or through advection from the corona. For clarity, we concentrated on the latter case, in which particles first cross a turbulent corona and
then enter a region of sheared velocity flow. There is a sizable uncertainty here, not only on the overall 
geometry of the problem, but also on the exact fraction of particles that are able to transit from one region to the 
next, as well as on the parameters characterizing the shear flow. We thus used the same parameters as in Sect.~\ref{sec:shear}  to describe the base of a mildly relativistic jet. To model the particle 
spectra, we convolved the Green function, which characterizes shear acceleration 
$G_{\rm shear}(\epsilon;\,\epsilon',\,\tau_{\rm shear})$, that is, the probability of jumping from $\epsilon'$ 
to $\epsilon$ over the timescale $\tau_{\rm shear}$ spent in the shear region, with an injection distribution 
that describes the time-dependent particle spectrum of the turbulent region. Denoting the latter 
as ${\rm d} n_{\rm turb}/{\rm d}\epsilon$, we write the stationary spectrum in the shear region as
\begin{equation}
    \frac{{\rm d}n_{\rm shear}}{{\rm d}\epsilon}\,=\,\frac{1}{\tau_{\rm adv,z}}
    \int_{0}^{\tau_{\rm adv,z}}{\rm d}\tau\,\int{\rm d}\epsilon'\,G_{\rm shear}(\epsilon;\,\epsilon',\,\tau)\,
    \frac{v_{\rm adv}}{v_{\rm shear}}\,\frac{{\rm d} n_{\rm turb}}{{\rm d}\epsilon'}\,,
    \label{eq:spec-shear}
\end{equation}
with $\tau_{{\rm adv},z}$ the advection time in the sheared region, as before, and
\begin{equation}
    \frac{{\rm d} n_{\rm turb}}{{\rm d}\epsilon'}\,=\,\int {\rm d}\epsilon''\,
    G_{\rm turb}(\epsilon';\,\epsilon'',\,\tau_{\rm adv})\,\frac{{\rm d} n_{\rm inj}}{{\rm d}\epsilon''}\,,
    \label{eq:spec-turb-out}
\end{equation}
where ${\rm d} n_{\rm inj}/{\rm d}\epsilon''$ characterizes (Sect.~\ref{sec:turb}) the injection spectrum 
at the entry of the turbulent region, as before. We introduced a ratio $v_{\rm adv}/v_{\rm shear}$ as a prefactor in 
Eq.~(\ref{eq:spec-shear}) to account for the difference in advection velocity of each region (conservation of particle current). 
We assumed that the shear region extends over $r_{\rm co}$ and took $\langle u_z\rangle=\beta_0\,c/2\simeq 0.5\,c$.

Finally, the total stationary spectrum was obtained as the sum of the spectra in each region, ${\rm d}n_{\rm cor}/{\rm d}\epsilon$ 
[Eq.~(\ref{eq:green-stoch})] and ${\rm d}n_{\rm shear}/{\rm d}\epsilon$ [Eq.~(\ref{eq:spec-shear})]. Although the overall 
normalization was left here as a free parameter, the relative normalization of these two was fixed by the volume of each region. 
The result is shown in the middle panel of Fig.~\ref{fig:combined}. This figure illustrates that the shear contribution acts as re-energization of the seed population produced by turbulent 
acceleration. This shapes an effective power-law tail close to $\propto \epsilon^{-3}$ (per number), which is close to the tail 
inferred from IceCube observations. The parameters we adopted are the same as in 
Sect.~\ref{sec:shear} for the shear layer, and $v_{\rm A}=0.20\,c$, $v_{\rm adv} = 0.02\,c$, 
$\ell_{\rm c}=10\,r_{\rm g}$, and $r_{\rm co}=30\,r_{\rm g}$ as in Fig.~\ref{fig:spec_stoch1} for the stochastic acceleration. 
For simplicity, we ignored possible boosting effects for the shear contribution. In general, if the shear flow were 
subrelativistic, shear acceleration would become too slow to have any effect, as quantified by the dependence of the 
diffusion coefficient on the Lorentz factor of the flow (Sect.~\ref{sec:shear}). We also note that we considered 
a scenario in which particles were injected at one boundary of the corona, accelerated by turbulence, and then advected 
into the shear layer. If we had considered that a fraction of particles could be advected into the black hole instead of 
the jet, the shear contribution would have been reduced. Similarly, if we had assumed that particles were injected into 
stochastic acceleration uniformly throughout the corona (see Sect.~\ref{sec:turb}, then the spectrum injected into the 
shear process would have been the steady-state spectrum ${\rm d}n_{\rm cor}/{\rm d}\epsilon$ instead of the spectrum discussed 
above. The overall shear contribution would have been effectively reduced in this case as well, and, a normalization of the 
overall spectrum to the inferred values without invoking boosting would have implied stronger energy 
requirements. In our case, the ratios of the integrated energy density in suprathermal particles to background 
plasma pressure are 4.1 and 2.7 for models (1) and (2), respectively.

Finally, we investigated the consequences of self-regulation of the acceleration process by turbulent damping. Because in the two scenarios discussed previously, the pressure carried by the accelerated particles is comparable to that 
in the background plasma, this last scenario appears to be quite likely. We also note that in the above cases, the flux 
normalization was set in such a way as to reproduce the normalization inferred from IceCube observations, and as mentioned 
previously, this amounts to selecting a precise fraction of particles that are extracted from the thermal pool are and 
then accelerated to high energies. We thus accounted for the backreaction of the accelerated particles on the 
turbulence following the prescriptions of \cite{2024PhRvD.109f3006L}. As detailed therein, backreaction becomes 
effective when the rate at which particles draw energy from the turbulent cascade exceeds the energy at which the cascade 
is replenished by energy injection. The former rate can be written $\sim \nu_{\rm acc}u_{\rm p}$ ($u_{\rm p}$ is the 
suprathermal proton energy density), with $\nu_{\rm acc} \sim v_{\rm A}^2/(c\ell_{\rm c})$, while the latter reads
$\sim (v_{\rm A}/\ell_{\rm c})\,u_B$ ( $u_B$ is the magnetic energy density). Turbulent damping therefore becomes effective 
when $u_{\rm p}/u_B \sim c/v_{\rm A}$. When we assume a plasma $\beta_p$ unity (see App.~\ref{sec:app}), this also means that 
damping takes place when $u_{\rm p}/p_{\rm gas} \sim c/v_{\rm A} \sim \mathcal O(1)$ here. When the particles start to damp 
the turbulence, the stochastic acceleration rate stalls, and acceleration becomes self-regulated. 

To fix the parameters, we assumed that the cascade rate on the outer scale was $0.5\,v_{\rm A}/\ell_{\rm c}$, as expected on 
general grounds. We used $r_{\rm co}=30\,r_{\rm g}$, $v_{\rm adv}=0.02\,c$, and $\ell_{\rm c}=10\,r_{\rm g}$ as before, and 
we injected particles at  $\gamma_0\sim 1$ as before, with an energy density lower by a factor of four than contained in the magnetized 
turbulence. We then tracked their acceleration using the two effective models (a) and (b) described by \cite{2024PhRvD.109f3006L}, 
which mimic models (1) and (2) above. In detail, model (a) corresponds to the Fokker-Planck model and assumed $v_{\rm A}=0.21\,c$, 
while model (b) describes advection in momentum space of a broken power-law spectrum assuming $v_{\rm A}=0.28\,c$. Finally, we 
computed the global stationary spectrum over the corona by integrating the time-dependent spectra as in Eq.~(\ref{eq:green-stoch}). 
The resulting spectrum is shown in the lower panel of Fig.~\ref{fig:combined}. We stress that in this scenario, no 
ad hoc normalization of the energy content was adopted. This energy content is dictated by the backreaction of particles 
on the turbulence, which, as discussed above, implies that the total energy density in suprathermal particles must be about 
$p_{\rm gas}c/v_{\rm A}$. In units of $p_{\rm gas}$, the energy density associated with the particle spectra shown in 
Fig.~\ref{fig:combined} is $3.4$ in model (a) and $4.1$ in model (b). The general shape of the spectrum does not depend much 
on the method used to carry out the integration. The key parameter that determines whether  the overall flux can match the detailed 
spectrum at energies $30-300\,$TeV is the Alfvén velocity (more generally, $\nu_{\rm acc}\tau_{\rm adv}$), which determines the 
location of the high-energy cutoff. The fact that the inferred spectrum can be reproduced at the price of reasonable parameters 
without an arbitrary normalization provides additional support and motivation for this scenario.

The global spectral shape of self-regulated acceleration can be understood as follows. In a first stage, particles are 
accelerated as in the test-particle picture of the first scenario that we investigated previously in this section. This occurs 
as long as the energy carried by the suprathermal particles remains below the threshold for backreaction, and it explains 
the similarity of the spectral slopes at low energies with that shown in the upper panel of Fig.~\ref{fig:combined}. When 
feedback becomes efficient, acceleration stalls at the momentum determining the peak of the energy distribution, while 
particles with a higher momentum can continue to be accelerated because they interact with wavelength modes of larger 
scales that have not yet been damped. This tends to shape time-dependent energy spectra that are approximately 
flat at high energies. The average of these time-dependent spectra over the crossing time of the corona, 
accounting for energy and escape losses, then shapes the observed spectrum, similar to or slightly steeper than $-2$ in 
slope (per number).  One phenomenological implication of this modeling is that in order to match a spectrum steeper 
than $-2$ in slope, the spectrum should be close to the cutoff region, that is, it should reveal increasing curvature toward higher 
energies. For higher acceleration rates or higher $\nu_{\rm acc}\tau_{\rm adv}$, the maximum energy is pushed to higher 
values, and the spectrum in a given region therefore approaches $-2$ in slope.

We therefore recall the dependence of the cutoff location on $\nu_{\rm acc}\tau_{\rm adv}$. However, if this combination of parameters were inhomogeneous throughout the corona, for example, because the Alfvénic velocity itself 
varied in space (as proposed in the context of X-ray binaries; see~\citealt{2024NatCo..15.7026N}), the sharp high-energy 
cutoff would be replaced by a softer dependence associated with the tail of values of $\nu_{\rm acc}\tau_{\rm adv}$. A more general interpretation of the proton spectrum is therefore called for, in which the flux level is determined 
by backreaction as modeled above, while the slope of the spectrum in the region probed by IceCube is dictated by 
the distribution of the acceleration rates (and advection times) experienced by the particle population. We defer 
the detailed investigation of this scenario to future work. 

\section{Conclusions}\label{sec:concl}
We have investigated scenarios of stochastic particle acceleration in a turbulent magnetized 
black hole corona and in the sheared velocity flow at the base of the jet to examine to which degree they can 
account for the proton energies inferred from IceCube neutrino observations of the Seyfert prototype NGC~1068, namely 
$\epsilon \sim 30-300\,$TeV, and for the overall level of the energy density. To this end, we used 
recent results for stochastic particle acceleration in turbulence and paid particular attention to the 
transport equation, that is, we considered the influence of turbulent transport on escape losses and 
incorporated advection losses as a time limit on particle acceleration (Sect.~\ref{sec:turb}). In 
Sect.~\ref{sec:disc} we compared three characteristic scenarios to the proton spectrum inferred in 
the case of NGC~1068 to conclude the following.

In a first scenario, which portrayed particle acceleration in the test-particle limit in a turbulent corona, 
we confirmed previous results that suggested that a characteristic Alfvénic velocity $v_{\rm A}\simeq 0.1-0.2\,c$ 
for a coherence length $\ell_{\rm c}\sim 3-10\,r_{\rm g}$ can reproduce the general shape of the observed spectrum. However, this comes at the cost of an ad hoc normalization of the proton spectrum up to an 
appreciable fraction of the background plasma pressure, which in some cases exceeds unity. In terms of phenomenology, the spectrum is expected to reveal an increasingly strong curvature toward higher energies.

In a second scenario, we investigated the possibility that stochastic acceleration consumes less 
energy, but is completed by a phase of reacceleration in the sheared velocity flow at the base of a mildly 
relativistic jet. This flow, whether interpreted as an inner disk wind or
jet wall, was envisioned to be launched from small radii well inside the corona; beyond this, however, it 
does not favor any specific model for the corona geometry. While this scenario admittedly introduces new parameters into 
the problem, it shows that shear reacceleration can play a role and lead to a satisfactory fit to the observed 
spectrum for a reasonable choice of parameters. In particular, depending on the conditions for escape, the 
particle distribution might extend to even higher energies, as shown in Fig.~\ref{fig:combined}. In our 
context, the relative importance of shear acceleration depends on the magnitude of the shear, which 
determines the rate at which particles are re-energized in the jet boundary layer, just as it depends 
on the advection velocity in the corona. Similarly to the first scenario, the overall flux normalization 
was chosen freely and required that the energy density in high-energy protons lies close to the overall pressure of the
background plasma. 

This apparent coincidence is intriguing with respect to particle injection into the stochastic process, 
because it means that a specific fraction of particles must be extracted from the thermal pool that is smaller by some orders 
of magnitude than unity, yet such that after acceleration, it carries an amount of energy that is comparable 
to the pressure of their parent population. Based on the observation that the gas pressure is $\beta_p$  
times the magnetic energy density $u_B$, and that $\beta_p\sim 1$, this observation may indicate that the acceleration
process is as efficient as can be, that is, that the high-energy particles extract as much energy as they can 
from the turbulent cascade. 

We therefore examined a third scenario in which turbulent acceleration was self-regulated by this feedback, 
which takes the form of damping of the turbulent cascade  in the inertial range. In this scenario, the problem was fixed 
by the rate at which turbulent energy is injected on the outer scale $\ell_{\rm c}$, with a generic value of 
$\sim u_B v_{\rm A}/\ell_{\rm c}$. Adopting this rate and characteristic Alfvénic velocities as above, 
solving for the concomitant dynamical evolution of the turbulence and of the accelerated particles, 
we were able to reproduce the inferred proton spectra without having to specify the flux normalization. 
We did not explore the possibility that the accelerated particles might provide a backreaction at a 
more global (hydrodynamical) level on the structure of the flow itself, such as the liftoff of 
the hydrostatic equilibrium in the corona or jet deceleration in the case of shear acceleration. These 
issues clearly deserve further investigation.  

This self-regulated acceleration process becomes operative provided the 
fraction $\chi_0$ of particles that are extracted from the thermal pool and injected into stochastic 
acceleration verifies $\chi_0\gtrsim (v_{\rm A}/c)^{-1} \beta_p^{-1}\,(\epsilon_{\rm max}/\overline 
\epsilon_{\rm th})^{-1}$ in terms of the plasma $\beta$ parameter ($\beta_p$), the Alfvénic velocity $v_{\rm A}$ 
(here playing the role of the characteristic eddy velocity on the outer scale of the turbulence), 
$\epsilon_{\rm max}$ the maximum energy of the nonthermal energy spectrum, and $\overline\epsilon_{\rm th}
\sim k_B T$ the mean thermal energy. If the injection fraction $\chi_0$ is smaller than the above, 
feedback has not set in by the time the particles are accelerated to $\epsilon_{\rm max}$, and hence,  particle 
acceleration proceeds as in a test-particle picture. However, if $\epsilon_{\rm max}/\overline\epsilon_{\rm th}\gg1$, 
that is, if particles are able to reach high energies, feedback becomes natural, and the high-energy 
part of the accelerated spectrum then depends weakly on the details of the injection. 

In this scenario of self-regulated stochastic acceleration, the source outputs as much energy as available in high-energy protons and achieves approximate equipartition between the supra-thermal proton and turbulent magnetic energy densities. This feature is all the more appealing in the context of high apparent neutrino luminosities such as were reported for NGC~1068. One direct consequence of this is that the overall neutrino luminosity of the source is not only controlled by the X-ray luminosity, which governs $p-\gamma$ interactions, but also by the magnetic energy content. Our study indicates that other parameters are likely to influence the neutrino spectrum. We notably remarked that the maximum energy is strongly sensitive to the acceleration rate, for example, to the 
Alfvén velocity, the coherence length, and to the advection timescale through the corona. This 
dependence arises because the mean particle energy increases exponentially fast in turbulent acceleration 
up to a number of e-folds that is set by a combination of these parameters. An important corollary of 
this is that different sources might exhibit different spectra. An alternative way of viewing this 
strong dependence is that if the corona were inhomogeneous in terms of $v_{\rm A}$ (e.g.), 
the final spectrum is a sum of spectra obtained for differing $v_{\rm A}$, or in other words, 
the final spectrum is shaped by the distribution of $v_{\rm A}$ (and other parameters) in the 
corona. Overall, it is therefore tempting to interpret the flux level as the consequence of 
self-regulated particle acceleration by turbulence damping (or possibly, viscous damping of the 
sheared velocity flow), and the general slope of the spectrum as given by the distribution of 
acceleration rates within the corona. 

Finally, as pointed out to us by A. Beloborodov and A. Levinson, the possibility of turbulence damping introduces new issues. In particular, it may jeopardize the energization of 
electrons, which is expected to power the X-ray luminosity of the corona. We note, however, that in 
our description, the advection of particles through the corona effectively leads to 
stratification in terms of mean energy because time-dependent spectra also depend on the position. 
Close to the boundary where particles are injected into the corona, turbulence is therefore not affected, and electron energization is efficient. Furthermore,
the damping of turbulence on scales of $\lesssim l$, for instance, does not prevent the formation of reconnection layers on scales $\sim l$, which offer ancillary energization channels. In an electron-ion plasma, reconnection can energize particles to a mean energy $\sim (1+ \sigma)m_p c^2$, where the magnetization parameter $\sigma\simeq (v_{\rm A}/c)^2$ (at $\sigma \lesssim 1$) denotes the ratio of the magnetic to plasma energy density~\citep{2018MNRAS.473.4840W}. While turbulent reconnection is therefore not expected to affect the proton spectrum at the highest energies for moderate values of $\sigma$, it may contribute to electron energization. In a conservative model, it suffices to energize electrons to Lorentz factors $\gamma_e< 10^2$ to power the X-ray luminosity, and these electrons behave 
as mildly or subrelativistic protons.  Overall, it appears reasonable to assume that the electrons  can indeed be energized as needed in spite of turbulent damping. A  detailed investigation of this question will form the basis of future work.

\begin{acknowledgements}
    It is a pleasure to thank A. Beloborodov and A. Levinson, F. Oikonomou and E. Peretti as well 
    as the members of the CN-6 connector of the Munich Excellence Cluster ORIGINS (in particular, 
    P. Padovani, X. Rodrigues, and E. Resconi) for stimulating discussions. FMR acknowledges 
    support by a DFG grant (RI 1187/8-1) and the kind hospitality of the IAP Paris.
    
\end{acknowledgements}

\bibliographystyle{aa} 
\bibliography{refs} 

\begin{appendix}
\section{Modelling the loss and escape rates}\label{sec:app}

\begin{figure}[h]
\centering
\includegraphics[width=0.48\textwidth]{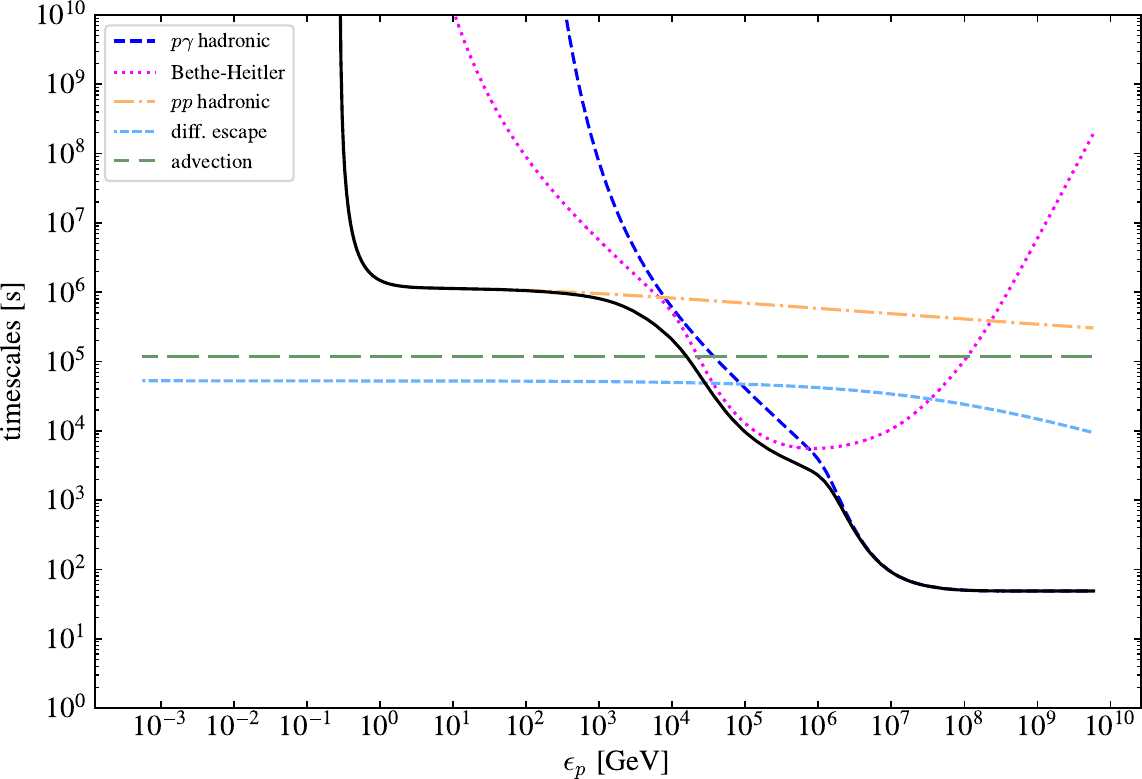}
\caption{Energy losses for: $L_{\rm disk}=5\times10^{44}\,$erg/s, $L_X=0.8\times 10^{44}\,$erg/s, $\Gamma_X=1.95$, $\epsilon_{X,\rm max}=128\,$keV, $r_{\rm co}=30\,r_g$, $r_g=2.35\times 10^{12}M_{7.2}\,$cm, $B=10^3\,$G, $n_p = 2\times 10^9\,$cm$^{-3}$. The diffusive escape time is shown for $v_{\rm A}=0.2\,c$, while the advection time assumes $v_{\rm adv}=0.02\,c$ (as expected for the radial inflow velocity at $r_{\rm co}$ with viscosity parameter $\alpha=0.1$).
\label{fig:eloss}}
\end{figure}
We incorporate energy and escape losses as follows. While the advection timescale $\tau_{\rm adv}$ is momentum-independent, $\tau_{\rm esc}\simeq r_{\rm co}^2/2 \kappa$ varies with momentum through the spatial diffusion coefficient $\kappa$. We model transport as a combination of turbulent advection with an effective diffusion coefficient $\kappa_{\rm turb}=\ell_{\rm c}v_{\rm A}/3$, and scattering on magnetic inhomogeneities with a mean free path $\lambda_{\rm scatt}=  r_{\rm L}^{1/3}\ell_{\rm c}^{2/3}$, see for example~\citet{1990acr..book.....B}, or \citet{2023JPlPh..89e1701L} and \cite{2023MNRAS.525.4985K} for recent discussions. Accordingly, we set $\kappa\equiv \kappa_{\rm turb} + \lambda_{\rm scatt}c/3$. As discussed in the main text, this indicates that particle escape is primarily governed by turbulent transport in the energy interval of phenomenological relevance, see Fig.~\ref{fig:eloss} for an illustration with our fiducial parameters $\ell_{\rm c}=10\,r_{\rm g}$, $v_{\rm A}=0.2\,c$, $r_{\rm co}=30 r_g$ and $B=10^3\,$G. Note that, when computing shear acceleration, we disregard the effect of turbulent transport, since we assume that the turbulence in the sheared region is not highly dynamic.

To model radiative losses, we consider a disk emission represented by a multicolor black-body 
$L_\epsilon\propto \epsilon^{4/3}\exp(-\epsilon/\epsilon_{\rm cd})$ for $\epsilon> 0.1 \epsilon_{\rm cd}$ and $L_\epsilon\propto 
\epsilon^3$ for $\epsilon<0.1 \epsilon_{\rm cd}$. The disk cutoff is $\epsilon_{\rm cd}=31.5\,$eV. The integrated disk 
luminosity is $L_{\rm disk}=5\times 10^{44}\,$erg/s. The X-ray emission is a powerlaw $L_\epsilon \propto 
\epsilon^{1-\Gamma_X}\exp(-\epsilon/\epsilon_{cX})$ for $\epsilon>\epsilon_{cd}$ (i.e., above the disk cutoff 
energy), and $L_\epsilon\propto \epsilon^2$ below. The X-ray cutoff is set at $\epsilon_{cX}=128\,$keV and 
the integrated luminosity $L_X=0.8\times 10^{44}\,$erg/s.

We assumed a black hole of mass $M_{\rm BH}=10^{7.2}\,M_\odot$, and corresponding gravitational 
radius $r_{\rm g}\simeq 2.35\times 10^{12}M_{7.2}\,$cm (Schwarzschild radius $r_{\rm s}=2r_g$). 
The Eddington reference luminosity is $L_{\rm Edd} = 2\times 10^{45} M_{7.2}$ erg/s. We note that
for NGC 1068, the all-flavour, integrated (1.5-15 TeV) isotropic-equivalent neutrino luminosity 
is of order $L_{\nu} \simeq 10^{42}(d/10~\mathrm{Mpc})^2$ erg/s $\sim L_{\rm Edd}/10^3$, and 
the estimated bolometric luminosity $L_{\rm bol} \sim 5\times 10^{44}$ erg/s \citep[e.g.,][]
{2022Sci...378..538I,Padovani2024}. We take $r_{\rm co}=30 r_g$ for the equatorial extension of 
the corona. The thermal background proton number density is set to $n_{\rm th}=2\times 
10^9\,$cm$^{-3}$, corresponding to a plasma $\beta$ parameter $\beta_p \sim 1$ (we use $1$ everywhere). 
Advective escape (radial inflow) in the corona  is parameterized by $v_{\rm adv}$, where we 
take $v_{\rm add} = \alpha_d v_k(r_{\rm co})$ as the reference value,  with $v_k(r)$ the Keplerian 
velocity field and $\alpha_d\simeq 0.1$ the viscosity parameter.

Figure~\ref{fig:eloss} shows the energy losses timescales for the given 
background photon spectral energy distribution. For comparison, for the chosen parameters, 
the required cosmic-ray power for a NGC~1068 type source is of order $L_{\rm CR} \sim 
L_{\nu} (c t_{\rm cool}/r_{\rm co}) \sim (5-10) \times 10^{43}$ erg/s, equivalent to an
energy density of $u_{\rm CR} \simeq L_{\rm CR}/(4\pi r_{\rm co}^2 c) \sim (3-6) \times 
10^4$ erg/cm$^3$ and comparable to the energy density $u_B\simeq 4\times 10^4$ erg/cm$^3$ 
of the magnetic field.

\end{appendix}

\end{document}